\documentstyle[aps,preprint]{revtex}
\begin{document}
\draft
\tighten
\preprint{\vbox{
\hfill  TIT/HEP-423/NP
 }}
\title{Use of the double dispersion relation in QCD sum rules 
with external fields}
\author{Hungchong Kim \footnote{E-mail : 
hckim@th.phys.titech.ac.jp, JSPS fellow}
}
\address{ Department of Physics, Tokyo Institute of Technology, Tokyo 
152-8551, 
Japan }

\maketitle
\begin{abstract}

In QCD sum rules with external fields, the double dispersion relation 
is often used to represent the correlation function.  
In this work, we point out that the double spectral density,
when it is determined by successive applications
of the Borel transformation, 
contains the spurious terms which should be kept in the
subtraction terms in the double dispersion relation.
They are zero under the Borel transformation but, if the
dispersion integral is restricted with QCD duality,
they contribute to the continuum. 
For the simple case with zero external momentum,  
it is shown that subtracting out the
spurious terms is equivalent to the QCD sum rules represented by
the single dispersion relation.

\end{abstract}

The QCD sum rule~\cite{SVZ} is widely used in studying hadronic
properties based on QCD~\cite{qsr}. In this framework, a correlation function
is introduced as a bridge between the hadronic and QCD representations.
In the QCD side, 
the perturbative part and the power corrections are calculated
in the deep space-like region ($q^2=-\infty$)
using the operator product expansion (OPE),
which is then used to extract the hadronic parameter of concern
by matching with the corresponding hadronic representation.

In matching the two representations, it is crucial to represent the
correlator using a dispersion relation.
Usually  in the nucleon mass sum rule as an example,
the single-variable dispersion relation is used. 
With this, the QCD correlator 
calculated in the deep space-like region can be related to its
imaginary part defined in the time-like region, which is then 
compared with the corresponding hadronic spectral
density to extract the hadronic parameter of concern. 
The hadronic spectral density contains contributions
from higher resonances as well as the pole from the low-lying resonance
of concern.    
To subtract out the continuum, QCD duality is invoked above a certain
threshold where the continuum contribution is equated to the
perturbative part of QCD.  This duality restricts 
the dispersion integral below the continuum
threshold in the matching.  Therefore, the predictive power of QCD sum
rules relies heavily on the duality assumption.  Indeed,
in the quantum mechanical examples, the parton-hadron duality 
works well for two-point correlation functions~\cite{blok}.

Often, within the QCD sum rule framework, a correlation function with 
an external field is considered
to calculate for examples pion-nucleon couplings~\cite{hat,hung,hung1}, 
nucleon magnetic moment~\cite{ioffe2}.
In such a case, 
as the two baryonic lines propagate through the correlator at the tree level,
the double-variable dispersion relation~\cite{ioffe} can be  
invoked to represent
the correlation function. Namely,
\begin{eqnarray}
\Pi (p^2_1, p^2_2) = \int^\infty_0 d s_1 \int^\infty_0  ds_2 { \rho (s_1, s_2)
\over (s_1 - p^2_1) (s_2 - p^2_2)} + {\rm subtractions}\ .
\label{double}
\end{eqnarray}
The subtraction terms serve to
eliminate infinities coming from the integral.
They are usually polynomials in $p^2_1$ or $p^2_2$, which vanish 
under the Borel transformations.  Thus, the subtraction terms
should not contribute to the sum rules.
As before, QCD duality is imposed to the correlator,
which restricts again the dispersion integral below a certain threshold
for both integration variables, $s_1$ and $s_2$.

In general, the double spectral density $\rho(s_1, s_2)$
in the double dispersion relation of Eq.~(\ref{double})
is obtained formally by matching the correlation function with
its corresponding OPE $\Pi^{\rm ope}$~\footnote{In this work, we focus mainly on
the OPE terms which contribute to the continuum.} calculated
in the deep Euclidean region using QCD degrees of freedom. 
That is,
\begin{eqnarray}
\Pi (p^2_1, p^2_2) = \Pi^{\rm ope} (p^2_1, p^2_2)+ {\rm subtractions}\ .
\label{sub}
\end{eqnarray}
The LHS contains the spectral density
in the integrals as given in Eq.~(\ref{double}).    
To solve for the spectral density $\rho(s_1, s_2)$ from this formal equation,
successive Borel transformations are needed to 
apply~\cite{beilin,braun}
on both side.  This will eliminate the unnecessary subtraction terms.
In doing so, the integrals are  
disappeared, reducing to a simple equation for the spectral density
in terms of a given OPE.
However, such a chosen spectral density, when it is put back to the
double dispersion integral Eq.~(\ref{double}), can reproduce the 
original OPE 
up to some {\it subtraction terms}.  Of course, the additional
subtraction terms do not matter if a
sum rule are constructed using precisely Eq.~(\ref{double}).
But in fact, in constructing continuum contribution, the duality 
argument is imposed, which further restricts the dispersion integral.   
If this restricted form of the dispersion integral is used, the subtraction 
terms do not vanish even after the Borel transformations.
In this work, we point out with some explicit examples that QCD sum 
rules using the double
dispersion relation contain these spurious contributions.

To proceed, we first demonstrate how the spectral density in the
double dispersion relation is usually determined~\cite{beilin,braun}.
The Borel transformation ${\cal B} (M^2, Q^2)$ is defined as
\begin{eqnarray}
{\cal B} (M^2, Q^2) f(Q^2)={\lim_{Q^2, n\rightarrow \infty ,~Q^2/n=M^2}}
{(Q^2)^{n+1} \over n !} \left (-{d\over dQ^2} \right )^n  f(Q^2)\ .
\end{eqnarray}
With this definition, the Borel transformation converts 
the $Q^2$ dependence of the function $f$ into
the Borel mass dependence, $M^2$.  
In doing so, polynomials in $Q^2$ vanish.
By applying the double Borel transformations on Eq.~(\ref{double}), 
we obtain,
\begin{eqnarray}
{\cal B} (M_2^2, -p^2_2){\cal B} (M_1^2, -p^2_1) \Pi (p^2_1, p^2_2)
=\int^\infty_0 ds_1 \int^\infty_0 
ds_2 \rho(s_1,s_2)~e^{-s_1/M^2_1-s_2/M^2_2}\ ,
\label{borelsum}
\end{eqnarray}
where we have used the formula,
\begin{eqnarray}
{\cal B} (M^2, Q^2) \left ( 1 \over Q^2 + \mu^2 \right)^n
= {1\over \Gamma(n) (M^2)^{n-1}} e^{-\mu^2/M^2}\ .
\label{bor1}
\end{eqnarray}
To eliminate the integral, we further perform additional
double Borel transformations and obtain,
\begin{eqnarray}
{\cal B} (\tau_2^2, {1\over M_2^2}){\cal B} (\tau_1^2, {1\over M_1^2}) 
{\cal B} (M_2^2, -p_2^2){\cal B} (M_1^2,-p_1^2) \Pi (p^2_1, p^2_2)
=\rho ({1\over \tau_1^2}, {1 \over \tau_2^2})\ ,
\label{spec}
\end{eqnarray}
where another formula  for the Borel transformation,
\begin{eqnarray}
{\cal B} (M^2, Q^2) e^{-a^2 Q^2}
= M^2 \delta (a^2 M^2-1)\ ,
\label{bor2}
\end{eqnarray}
has been used.  Note in this derivation, 
$s_1=1/\tau^2_1 \ge 0$ and  $s_2=1/\tau^2_2 \ge 0$, thus restricting
the spectral density only in the region $s_1, s_2 \ge 0$.

The OPE spectral density can be obtained by applying this
operation on a given OPE.  
Note also that the integral interval should include the
point provided by the delta functions.  If the interval  does not
include the point, $s_1=1/\tau^2_1$ or $s_2=1/\tau^2_2$, then
Eq.~(\ref{spec}) is not conclusive.
We stress that  
the spectral density determined via Eq.~(\ref{spec}) is correct
within this context, obtained by successive applications of Borel 
transformation
on the dispersion integral limited from zero to infinity.
The subtraction terms, as they vanish under the Borel transformations,
can be chosen freely as we like.

In practice, however,  QCD sum rules require a certain assumption
for high energy part of the correlator, QCD duality.
With this assumption, the dispersion integral 
is restricted below the continuum threshold $S_0$, and the
Borel-transformed sum rule becomes,
\begin{eqnarray}
\int^{S_0}_0 ds_1 \int^{S_0}_0 ds_2 \rho^{\rm ope}
(s_1,s_2)~e^{-s_1/M^2_1-s_2/M^2_2}
=
\int^{S_0}_0 ds_1 \int^{S_0}_0 ds_2 \rho^{\rm phen}
(s_1,s_2)~e^{-s_1/M^2_1-s_2/M^2_2}
\label{bsum}
\end{eqnarray}
$\rho^{\rm phen} (s_1,s_2)$ is obtained  from hadronic representation of the
correlator while
$\rho^{\rm ope} (s_1,s_2)$ is obtained via
Eq.~(\ref{spec}) for a given OPE.
The LHS restricted below
the continuum threshold $S_0$ can be calculated directly
using the spectral density obtained from Eq.(\ref{spec}).
Another equivalent method which is more useful for our discussion 
is to calculate the LHS via
\begin{eqnarray}
{\cal B} (M_2^2, -p_2^2){\cal B} (M_1^2,-p_1^2) \left [
\Pi^{\rm ope} (p^2_1, p^2_2)
-\int^\infty_{S_0} ds_1 \int^\infty_{S_0} ds_2 {\rho^{\rm ope} (s_1, s_2)\over
(s_1-p_1^2)(s_2-p_2^2) } \right ]\ .
\label{bope}
\end{eqnarray}
Note that the continuum is subtracted out from the OPE using the
duality argument.  Integral intervals like $\int^\infty_{S_0} ds_1
\int^{S_0}_0 ds_2$ or $\int^{S_0}_0 ds_1 \int^\infty_{S_0} ds_2$
do not contribute because, as we will see,
the spectral density is of the form $\sim \delta(s_1-s_2)$ at least
in our examples that will be considered in this work.
The integral in the second term is bounded below by $S_0$.
As we have stressed above,  in determining the spectral density 
via Eq.~(\ref{spec}),  it is important that the dispersion integral is
limited from zero to infinity.
Since the second integral is bounded below by $S_0$ due to the
duality argument, it is not clear
if the subtraction terms as written in Eq.~(\ref{double})
do not participate in the sum rule.
This is our main question to be addressed in this work.
 
Let us proceed how our question is realized in
QCD sum rules with external fields.
To do so, we consider as an example
the two-point correlation function with a pion,
\begin{eqnarray}
\Pi (q, p_\pi) = i \int d^4 x e^{i q \cdot x} \langle 0 | T[J_N (x)
{\bar J}_N (0)]| \pi (p_\pi) \rangle \ ,
\label{two}
\end{eqnarray}
where $J_N $ is the nucleon interpolating field 
proposed by Ioffe~\cite{ioffe3}.
To be specific, let us take a typical OPE from this correlation function,
\begin{eqnarray}
\Pi^{\rm ope}_1=\int^1_0~du~\varphi_p (u)~ln[-(q-up_\pi)^2]
\sim \int^1_0~du~\varphi_p (u)~ln[-up^2_2-(1-u)p^2_1]\ .
\label{ope1}
\end{eqnarray}
Here $p^2_1 = q^2$ and $p^2_2=(q-p_\pi)^2$.
We have taken the limit $p_\pi^2=m_\pi^2=0$ as is usually done in the
light-cone QCD sum rules~\cite{bely}.  
Note, Eq.~(\ref{ope1}) contains terms polynomials in $p_1^2$ or
$p_2^2$ but we did not specify these subtraction terms explicitly.  
For the twist-3
pion wave function, we take its asymptotic form,
$\varphi_p (u)=1$~\cite{bely}.  With 
higher conformal spin
operators, the wave function takes more complicate form but our claims
in this work are still valid even with more general wave functions. 
We will discuss this point later.

To obtain the double spectral density,
we take the operation as given in Eq.~(\ref{spec}).
For $\Pi^{\rm ope}_1$, we straightforwardly obtain
\begin{eqnarray}
\rho^{\rm ope}_1 (s_1, s_2) = - s_1 \delta (s_1- s_2)\ .
\label{opespec}
\end{eqnarray}
Note, the spectral density is defined only in the
region $s_1, s_2 \ge 0$.  Therefore, the spectral density should
be understood as being multiplied by the step function,
$\theta(s_1)\theta(s_2)$.  
To include entire region of $\rho(s_1, s_2) \ne 0$, 
the lower boundary of the dispersive integral should be understood 
as $0^-$, infinitesimal but negative value.   
This subtlety does not matter in this example but is important in later
examples.

Normally, $\rho^{\rm ope}_1$ is simply used in QCD sum rules
Eq.~(\ref{bope}) without justifying its use carefully. 
To see a problem with this spectral density, 
we put this expression into Eq.~(\ref{double}) and
perform 
the integrations using the Feynman parametrization,
\begin{eqnarray}
&&\int^\infty_0 ds_1 \int^\infty_0 ds_2 {\rho^{\rm ope}_1 (s_1,s_2) \over 
(s_1-p_1^2) (s_2 -p_2^2) }=
\int^1_0 du \int^\infty_0 ds {- s \over [s - up_2^2 -(1-u)p^2_1]^2}\nonumber \\
&&
= - \int^1_0 d u \left [
{ -up_2^2 - (1-u) p_1^2 \over s -up_2^2 - (1-u) p_1^2} \Bigg |^\infty_0
+ln[s-up_2^2 - (1-u) p_1^2] \Bigg |^\infty_0
\right ] \ .
\label{spur}
\end{eqnarray}
The second term in the last line is the anticipated logarithmic term matching
the OPE of Eq.~(\ref{ope1}). In other words, the second term is enough
to reproduce the Borel-transformed OPE of Eq.~(\ref{ope1}).
This means that the
first term is spurious and it vanishes under Borel transformations
with respect to the variables, $-p^2_1$ and $-p^2_2$.
Therefore,  it is a part of subtraction terms and should not
contribute to the QCD sum rule.  That is, we have to subtract out this
term using our freedom to choose any subtraction term. 
Of course, this subtraction by hand is not necessary if the
sum rule is used in the context of Eq.~(\ref{borelsum}).
However, in practice, the sum rule is used in the context of Eq.~(\ref{bope})
invoking QCD duality.
The continuum part from this subtraction term,
\begin{eqnarray}
- \int^1_0 d u 
{ -up_2^2 - (1-u) p_1^2 \over s -up_2^2 - (1-u) p_1^2} \Bigg |^\infty_{S_0}\ ,
\end{eqnarray}
becomes, under the double Borel transformation,
\begin{eqnarray}
S_0 M^2 e^{-S_0/M^2}~~{\rm where}~~ {1\over M^2} = {1\over M_1^2}
+ {1\over M_2^2}\ .
\end{eqnarray}
This nonzero continuum is spurious as it originated from 
the subtraction term.  A prescription for including the continuum
presented in Ref.~\cite{bal} can be obtained when this spurious continuum
is kept and it is often used in the light-cone QCD sum rules~\cite{zhu}.    
However, keeping this term while neglecting the
OPE subtraction terms in Eq.~(\ref{ope1}) is inconsistent.

We now consider a slightly different OPE from Eq.~(\ref{two}),
\begin{eqnarray}
\Pi^{\rm ope}_2 = \int^1_0 du \varphi_p (u) {1\over -up^2_2-(1-u)p^2_1} 
\label{ope2}
\end{eqnarray}
Again we take the asymptotic form for the pion wave function $\varphi_p (u)=1$
for simplicity.  Note, when the external momentum is zero, we have 
$\Pi^{\rm ope}_2 = - 1/p^2~~ (p^2_1=p^2_2= p^2)$.  It is clear that
this OPE 
should not contribute to the continuum.  Even if the sum rule
is constructed with nonzero external momentum, this aspect should
be recovered whenever we take the external momentum zero. 
Under the successive applications of the Borel transformation,
it is straightforward to obtain the corresponding
spectral density,
\begin{eqnarray}
\rho^{\rm ope}_2 (s_1, s_2) = \delta(s_1-s_2) \theta(s_1)\ . 
\end{eqnarray}
Here we put the step function explicitly because the subtlety
associated with the lower boundary affects the discussion.
Using this spectral density, 
Eq.~(\ref{double}) becomes
\begin{eqnarray}
\int^1_{0} du \int^\infty_{0^-} ds 
{\theta(s) \over [s - up_2^2 -(1-u)p^2_1]^2}\ .
\end{eqnarray}
Note, we have $0^-$ for the lower limit of the integral in order to ensure
that the integration includes entire region of $\rho_2 (s_1, s_2)\ne 0$. 
Integration by part leads to
\begin{eqnarray}
\int^1_0 du \int^\infty_{0^-} ds 
{\delta(s) \over s - up_2^2 -(1-u)p^2_1} 
-\int^1_0 du { \theta(s) \over s-up_2^2 -(1-u)p^2_1} \Bigg |^{\infty}_{0^-}\ .
\label{www}
\end{eqnarray}
Here the first term yields the anticipated OPE of Eq.~(\ref{ope2}) and
the second term, as the lower limit lies just below the zero, is
zero.  
Note also that this separation becomes possible
because the subtlety with the lower boundary.  If there were no
subtlety with the lower boundary, then we would not have
the first term containing the delta function.  
If this were true, then in the limit of zero external
momentum, Eq.(\ref{ope2}) can not be equal to Eq.(\ref{www}),
which does not make sense.

It is the second term that should be a part of the subtraction
terms.  Of course, this separation does not have 
any mathematical significance.
It is however important physically because this separation
enables us to identify where the spurious contribution to
the sum rule comes from.  That is, the second
term, when the lower limit changes to $S_0$, survives
under the Borel transformations and contributes to the continuum. 
Once again, we have identified a spurious subtraction
which contributes to the sum rule.

Up to now, from the two simple examples,
we have shown that QCD sum rules invoking the double
dispersion relation contain the spurious terms originated from
the subtraction terms.  
They contribute to the sum rule when QCD duality is imposed.
This is a general statement as long as
QCD sum rules are constructed using Eqs.~(\ref{bsum}),(\ref{bope}) while
the spectral density is determined via Eq.~(\ref{spec}). 
In general, for the correlator of Eq.~(\ref{two}) as an example,
the OPE contains complicate
functions like the pion wave functions. 
The general twist-3 pion wave function can be written~\cite{bely},
\begin{eqnarray}
\varphi_p (u) = \sum_k a_k u^k\ .
\end{eqnarray}
Using this general form, $\Pi^{\rm ope}_2$ under the 
double Borel transformations becomes 
\begin{eqnarray}
\varphi_p (u_0) M^2
\end{eqnarray}
where $u_0 = M_1^2/(M_1^2+M_2^2)$ and $M^2 = M_1^2 M_2^2/(M_1^2+M_2^2)$.
Under additional Borel transformations, the spectral density
can be shown to be proportional to derivatives of $\delta (s_1-s_2)$.
The dispersion integral of Eq.~(\ref{double}) can be
performed by integrations by part but, in doing so, 
the boundary terms become parts of the subtraction terms and
produce the spurious continuum contributions when the sum rule
is combined with QCD duality.
In general, it is difficult to eliminate these
spurious terms systematically.  
This is  a generic problem in the sum rules using
the double dispersion relation combined with QCD duality.

Now, let us consider  a simple case with the zero external momentum. 
As a specific example, we consider the
two-point correlation function with a pion with the vanishing pion momentum
(the soft-pion limit)
or the same correlation function with one pion momentum taken out as 
an overall factor but in the rest with $p_\pi^\mu =0$~\cite{hung,hung1,krippa}
(beyond the soft-pion limit).
Even in this case, the double dispersion relation is proposed
as a correct representation of the correlator~\cite{ioffe}.
The double dispersion relation might be useful in treating 
phenomenological side properly but the spurious terms
still persist.

The double spectral density in this case takes the form 
\begin{eqnarray}
\rho (s_1, s_2) = \rho (s_1) \delta (s_1-s_2)\ .
\end{eqnarray}
Since the two correlator momenta are equal in this case, 
the delta function appears as a part of the spectral
density. 
The double dispersion relation Eq.~(\ref{double}) reduces to
\begin{eqnarray}
\Pi (p^2) = \int^\infty_0 ds {\rho (s) \over (s-p^2)^2} + 
{\rm subtractions}\ .
\end{eqnarray}
Unlike to the single dispersion relation, the correlation function
contains square of $s-p^2$ in the denominator.  
The spectral density $\rho(s)$ is obtained by
\begin{eqnarray}
{\cal B} (\tau^2, {1\over M^2}){\cal B} (M^2, -p^2) \Pi (p^2)
= {\partial \rho (s) \over \partial s} \Bigg |_{s=1/\tau^2}
\label{sing}
\end{eqnarray}
Thus, we can determine the derivative of the spectral density
for a given OPE.
The OPE corresponding to Eq.~(\ref{ope1}) is
$ln(-p^2)$. We do not need to 
worry about
the pion wave function $\varphi_p (u)$ since its overall normalization,
which is fixed to the unity,
participates in the case.

Substituting $ln(-p^2)$ into Eq.~(\ref{sing}), 
we obtain
\begin{eqnarray}
{\partial \rho (s) \over \partial s} \Bigg |_{s=1/\tau^2} = -1~~
\Longrightarrow~~ \rho (s) =-s + {\rm constant} \ .
\label{sope}
\end{eqnarray}
The constant term, when put into the dispersion integral, yields
the term $1/p^2$.  To be consistent with the logarithmic
behavior of the
OPE, the constant should be zero.
The rest of the spectral density leads to
the dispersion integral,
\begin{eqnarray}
\int^\infty_0 ds {-s \over (s-p^2)^2} = 
{p^2 \over s-p^2} \Bigg |^\infty_0 - ln(s-p^2) \Bigg |^\infty_0\ .
\label{sss}
\end{eqnarray}
The second term in the RHS is what we have anticipated.  This is what one
would have obtained if the single dispersion relation is used.
Again, this is enough to reproduce the Borel-transformed OPE of $ln(-p^2)$. 
But, as before, the first term in the RHS 
is spurious.  This term is zero under the Borel transformation but
the continuum gives nonzero contribution to the sum rule.
Note that the first term can be
separated as
\begin{eqnarray}
{p^2 \over s-p^2} \Bigg |^\infty_0 = {p^2 \over s-p^2} \Bigg |^{S_0}_0 +
{p^2 \over s-p^2} \Bigg |^\infty_{S_0}\ .
\end{eqnarray}
What is interesting is that the contribution from the upper limit in 
the first term cancels the one from the lower limit
in the second term.  These two terms coming from the continuum 
threshold survive separately under the Borel transform though
their sum is still zero.
If the spectral density of Eq.~(\ref{sope}) is simply used
in the sum rule of Eq.~(\ref{bsum}), then the lower limit from the
second term contributes to the continuum while the upper limit from the
first term does not participate to the sum rule Eq.~(\ref{bsum}). 
Once again, the spurious 
nature of this continuum is obvious.

Another way to support our claim is to 
consider the correlator Eq.~(\ref{two}) in the soft-pion limit.
According to the soft-pion theorem, the
correlator becomes the commutator with the
axial charge $Q_5$,
\begin{eqnarray}
\Pi (q,p_\pi=0) \sim 
i \int d^4 x e^{i q \cdot x} \langle 0 | \left [ Q_5, T[J_N (x)
{\bar J}_N (0)] \right ]|0 \rangle \ .
\end{eqnarray} 
The commutator, if it is readily evaluated using
the commutation relation for the quark fields, becomes the anticommutator,
\begin{eqnarray}
\Pi (q,p_\pi=0) 
\sim \left \{ \gamma_5, ~i \int d^4 x e^{i q \cdot x} \langle 0 | T
\left [ J_N (x)
{\bar J}_N (0) \right ] |0 \rangle \right \}\ .
\label{soft}
\end{eqnarray}
It means, in the soft-pion limit, Eq.~(\ref{two})
is equivalent to the nucleon chiral-odd sum rule, which
should be represented by the single dispersion relation in the construction
of its sum rule.  Note, in deriving this, we have used only the
soft-pion theorem and the commutation relation for the quark fields.
Therefore, this is an operator identity that has to be satisfied always.
But if one starts from the double dispersion
relation and takes the soft-pion limit afterward, then 
Eq.~(\ref{soft}) is not satisfied exactly by the presence of the spurious term
like ${p^2 \over s-p^2} \Big |^\infty_0 $ in Eq.~(\ref{sss}).

Indeed, the spurious terms
lead to different continuum factors as appeared in
Ref.~\cite{krippa,aw}.  
A similar discussion can be found in 
Ref.~\cite{hung2} where we simply point out
a pole at the continuum threshold, not clarifying its spurious
nature.  It is currently under-debate whether or not keep the
terms in question in QCD sum rules with external fields\cite{bkr}.
But it is now clear from our discussion that they are spurious. 
Anyway, once the spurious subtractions are eliminated, then what is 
left is the
same sum rule that we would have obtained via the single dispersion 
relation.  
Our argument can be generally applied to other OPE contributing 
to the continuum and it can be shown that subtracting the
spurious continuum leads to the sum rule invoking the single
dispersion relation.  
Normally, the continuum contribution is
denoted by the factors, $E_n (x\equiv S_0/M^2) = 1 - 
(1 + x + \cdot\cdot\cdot + x^n/n!) e^{-x}$. 
Subtracting the spurious continuum replaces the continuum factor,
$E_n (x) \rightarrow E_{n-1} (x)$ for $n \ge 1$ .
The changes due to this spurious term are sometime
huge as discussed in Ref.~\cite{hung1}. In Ref.~\cite{hung1}, 
figure 1
shows for the sum rules with $i\gamma_5 p_\mu \gamma^\mu$ structure
how this spurious continuum changes the Borel curve of $\pi NN$ 
coupling.  There, the Borel curve with the spurious continuum varies 
within the range 11.5 - 11.2.  But without this spurious continuum,
the variation scale becomes 15 - 30, clearly
showing huge effects from the spurious terms. 
The $\pi NN$ coupling extracted from
this Borel curve will be quite different from what we know experimentally.
As discussed in Ref.~\cite{hung1}, however, this only says that
the Dirac structure $i\gamma_5 p_\mu \gamma^\mu$ is 
not adequate for calculating the coupling. Instead, $i\gamma_5$
or $\gamma_5 \sigma_{\mu\nu} q^\mu p^\nu$ Dirac structure is more
useful to calculate the coupling~\cite{hung}.  Nevertheless, from this 
consideration,  we see that
the spurious continuum sometime becomes substantial and
thus should be treated carefully.


In summary, we have pointed out in this work that 
QCD sum rules with external fields employing a double dispersion
relation can contain the
spurious continuum originated from the subtraction terms.
This spurious term
appears because the spectral density obtained via successive applications
of the Borel transformation is not compatible with QCD duality.
The spurious term should be subtracted out using the freedom
for subtraction terms in QCD sum rules.
In the case with the zero external momentum, subtracting the
spurious term is equivalent to the sum rules invoking
the single dispersion relation.
Of course, our finding only affects continuum contributions
but in some cases this modification leads to significant corrections
to QCD sum rule results as is presented in Ref.~\cite{hung1}.

\acknowledgments
This work is supported by Research Fellowships of
the Japan Society for the Promotion of Science.

\end{document}